\documentclass{aa}

\newcommand\simlt{\lower.5ex\hbox{$\; \buildrel < \over \sim \;$}}
\newcommand{\fermi}{{\it Fermi}-LAT\xspace}
\newcommand{\gray}{$\gamma$-ray\xspace}
\newcommand{\grays}{$\gamma$-rays\xspace}
\newcommand{\ct}{CTA 102\xspace}
\DeclareUnicodeCharacter{B0}{\textdegree}
\usepackage{amsmath}
\usepackage{graphicx}
\usepackage{longtable}
\usepackage{multirow}
\usepackage{lscape}
\usepackage[colorlinks=true, citecolor=blue, linkcolor=blue, urlcolor=blue]{hyperref}
\usepackage{txfonts}
\begin{document} 
   \title{Investigation of the Gamma-ray Spectrum of CTA 102 During the Exceptional Flaring State in 2016-2017}
   \author{N. Sahakyan
          \inst{1,2}
          }
   \institute{ICRANet, P.zza della Repubblica 10, 65122 Pescara, Italy.
         \and
             ICRANet-Armenia, Marshall Baghramian Avenue 24a, Yerevan 0019, Armenia.
             \email{narek@icra.it}
             }

   \date{Received -; accepted -}


  \abstract
  {The flat spectrum radio quasar CTA 102 entered an extended period of activity from 2016 to 2017 during which several strong $\gamma$-ray flares were observed. Using Fermi large area telescope data a detailed investigation of \gray spectra of CTA 102 during the flaring period is performed. In several periods the \gray spectrum is not consistent with a simple power-law, having a hard photon index with an index of $\sim(1.8-2.0)$ that shows a spectral cutoff around an observed photon energy of $\sim(9-16)$ GeV. The internal $\gamma$-ray absorption via photon-photon pair production on the broad-line-region-reflected photons cannot account for the observed cut-off/break even if the emitting region is very close to the central source. This cut-off/break is likely due to a similar intrinsic break in the energy distribution of emitting particles. The origin of the spectral break is investigated through the multiwavelength modeling of the spectral energy distribution, considering a different location for the emitting region. The observed X-ray and $\gamma$-ray data is modeled as inverse Compton scattering of synchrotron and/or external photons on the electron population that produce the radio-to-optical emission which allowed to constrain the power-law index and cut-off energy in the electron energy distribution. The obtained results are discussed in the context of a diffusive acceleration of electrons in the CTA 102 jet.}

   \maketitle
   
\section{Introduction}\label{sec1}
Jets are observed in many classes of astrophysical objects, ranging from active galactic nuclei (AGNs) to galactic binary systems \citep[e.g.,][]{2005AdSpR..35..908D,2017mbhe.confE..62B}. The jets mostly likely being powered by accretion processes are among the most powerful emitters of radiation in the Universe. Undoubtedly, AGNs are one of the most representative classes of astrophysical objects where the jets can be studied in all scales. The jets in these objects are manifestation of energy release from super-massive black holes (with masses up to $10^9M_\odot$) and they can extended to several hundreds of kiloparsecs and in some cases to a few megaparsecs into the space often remaining highly collimated. Now the emission from these jets can be observed not only form their innermost regions, where it is stronger, but also from extended components. For example, the emission up to the X-ray band from the extended knots or hot spots of relativistic jets are observed \citep[e.g.,][]{2002ApJ...565..244H,2004ApJ...608...95S, 2005ApJ...622..797K}, the origin of the emission being explained either by synchrotron emission \citep{2006ARA&A..44..463H} or by the inverse Compton (IC) scattering of either synchrotron photons \citep{hardcastle2002} or cosmic microwave background (CMB) photons \citep{2000ApJ...544L..23T, 2001MNRAS.321L...1C, 2017A&A...608A..37Z}. However, in some cases, the IC/CMB model has been ruled out \citep[see][]{2014ApJ...780L..27M,2015ApJ...805..154M,2017ApJ...835L..35M,2017ApJ...849...95B}. Several alternative emission models for the knots involve the radiation of protons \citep[e.g.,][]{2016ApJ...817..121B, 2017ApJ...835...20K, 2002MNRAS.332..215A}. Even though the observations of knots, hot-spots and lobes caries significant information on the jet energetics and dynamics but to understand the central source and the formation and propagation of the jets, it is necessary to carry out extensive studies of their initial sub-parsec-scale region.\\
Observation of blazars is the best way to explore the physics of jets. Blazars are a subclass of AGNs with a dominant nonthermal emission from a jet that is closely aligned with the observer's line of the sight \citep{urry}. Such a geometry leads to the relativistic Doppler amplification of the emission and the radiation appears brighter for the observer. Because of this the blazars are observed even at very high redshifts \citep[e.g.,][]{2017ApJ...837L...5A}. Blazars are the most luminous and energetic objects in the known universe and are the dominant sources in the extragalactic \gray sky. In the High Energy (HE; $>100$ MeV) \gray band, among the 5000 sources detected to date more than 3100 are blazars \citep{2019arXiv190210045T}. One of the most distinct features of blazars is rapid variability across the whole electromagnetic spectrum with the most dramatic and short time scale changes being observed in the \gray band \citep[e.g., minute scales,][]{2016ApJ,foschini11,foschini13, nalewajko, brown13, rani13,saito,hayashida15}. This strongly constrains the emitting region size (by the light travel considerations), suggesting the radiation comes from a compact region of the jet. Traditionally the blazars are classified based on their emission lines: flat-spectrum radio quasars (FSRQs) exhibit broad emission lines, while BL Lacs show weak or no emission lines in their optical spectra. A different classification is based on the synchrotron peak frequency ($\nu_{\rm p}$): when $\nu_{\rm p}$ is in the infrared, optical, or ultraviolet/X-ray bands low synchrotron peak (LSP), inter- mediate synchrotron peak (ISP), and high synchrotron peak (HSP) sources are classified respectively \citep{1995ApJ...444..567P, 2010ApJ...716...30A}. Typically FSRQs are LSP/ISP blazars, whereas, BL Lacs are mostly HSP ones.\\
The broadband spectral energy distribution (SED) of blazars exhibits a double peaked structure, one between the infrared and X-ray bands (low energy component) and the other above the X-ray band (HE component). It is well established that the low energy component is from the synchrotron emission of electrons in the magnetic field of the jet but the nature of HE component is less well understood. The HE component is most likely due to the IC up scattering of the low energy photons produced either inside (synchrotron self Compton (SSC), \citep[][]{ghisellini, bloom, maraschi}) or outside of the jet (external inverse Compton (EIC)) \citep{blazejowski,ghiselini09, sikora}). The nature of the external photon field depends on the location of the emitting region and can be either the photons directly emitted from the disk or those reflected from the broad line region (BLR) or infrared photons emitted from the dusty torus or photons from dusty torus clouds irradiated by a spine-sheath jet when the emission region is further ($>{\rm pc}$) from the central objects \citep{2018ApJ...853...19B}. SSC scenario was successfully applied to model the broadband SED of BL Lacs, while the SEDs of FSRQs are better explained by EIC models. Other possible processes used to model the SEDs of blazars invoke the acceleration and emission of jet accelerated protons. These models recently were more frequently applied to model multiwavelength and multimessneger observations of TXS 0506+056 - the first cosmic neutrino source \citep{2018ApJ...863L..10A,2019NatAs...3...88G, 2019MNRAS.483L..12C, 2018ApJ...864...84K, 2018ApJ...865..124M, 2018arXiv180705210L,2018arXiv180900601W,2018ApJ...866..109S} as well as its neighboring blazar PKS 0502+049 \citep{2019A&A...622A.144S}.\\
\ct is one of the bright blazars observed by Fermi Large Area Telescope (Fermi LAT) in the HE \gray band. Even with its large distance, $z=1.037$, \ct sometimes shows strong \gray outburst with a flux exceeding $10^{-5}\:{\rm photns\:cm^{-2}\:s^{-1}}$. For example, on 16 December 2016 (during the prolonged \gray activity) within 4.31 minutes the \gray flux above 100 MeV was as high as $(3.55\pm0.55)\times10^{-5}\:{\rm photns\:cm^{-2}\:s^{-1}}$, corresponding to an isotropic \gray luminosity of $ L_{\rm \gamma} =3.25\times 10^{50}\: {\rm erg\:s^{-1}}$, which is among the highest luminosities so far observed in the \gray band \citep{2018ApJ...863..114G}. In addition, \citet{2018ApJ...854L..26S} showed that the \gray flux variation time can be as short as $\sim5$ minutes. The analysis of multiwavelength light curves showed correlated variations in all the observed energy bands indicating co-spatial origin of the emissions \citep{2018A&A...617A..59K, 2018ApJ...863..114G}. The broadband emission of \ct is better modeled when the photons external to the jet (infrared photons from the torus) are considered \cite{2018ApJ...863..114G}. As an alternative interpretation, the ablation of a gas cloud penetrating the relativistic jet of \ct was discussed to be the source of the observed emission \citep{2017ApJ...851...72Z, 2019ApJ...871...19Z}.\\
The previous studies indicated a deviation of the \gray spectra of \ct from a power-law model at HEs \citep{2018A&A...617A..59K, 2018ApJ...863..114G}. Such breaks have already been observed in the \gray spectra of several blazars which can be of different origin, varying from source to source. In principle, if the emitting region is within the BLR sharp breaks in the the \gray spectra are expected to be due to strong attenuation of the HE and very high energy ($>100$ GeV; VHE)  photons through their interaction with the optical photons. The optical depth for the interaction of several tens of GeV photons can be very large, preventing their escape from the region. So, if the break is due to the absorption, it will put a constraint on the location of the \gray emitting region which is crucial when modeling of the observed data.  Of course, a possible break in the GeV spectra does not necessarily imply absorption due to BLR photons; such break can be also due to the underlying (radiating) electrons with energy distributions deviating from a power-law spectrum. Since the \gray emission is caused by the IC up-scattering of low energy photons, the shape of the \gray spectra is directly related to the energy distribution of accelerated electrons. Thus, the modeling of the \gray spectra with a break can allow to probe the highest tail of the energy distribution of underlying electrons which is formed in the interplay between the acceleration and cooling of the particles. So, this is a powerful tool for diagnosing the physics of particles in the jets. A curvature (break), if statistically significant, contains wealth of information on the possible location of the emitting region and/or on the acceleration and cooling of the particles. In this regard \ct is an ideal target considering the previous indication of deviation of its \gray spectrum from a power-law model and availability of a large amount of simultaneous multiwavelength data.\\
This paper is organized as follows. The \gray data analysis is presented in Section \ref{sec2}. The origin of the observed breaks is investigated in Section \ref{abs}. In Section \ref{sec4}, the origin of multiwavelength emission is discussed for a different location of the emission region. The formation of the energy distribution of radiating electrons taking into account their acceleration and cooling is investigated in Section \ref{sec5}. The results are presented and discussed in Section \ref{sec6} while the conclusion is summarized in Section \ref{sec7}.
\begin{figure*}
   \includegraphics[width=0.99 \textwidth]{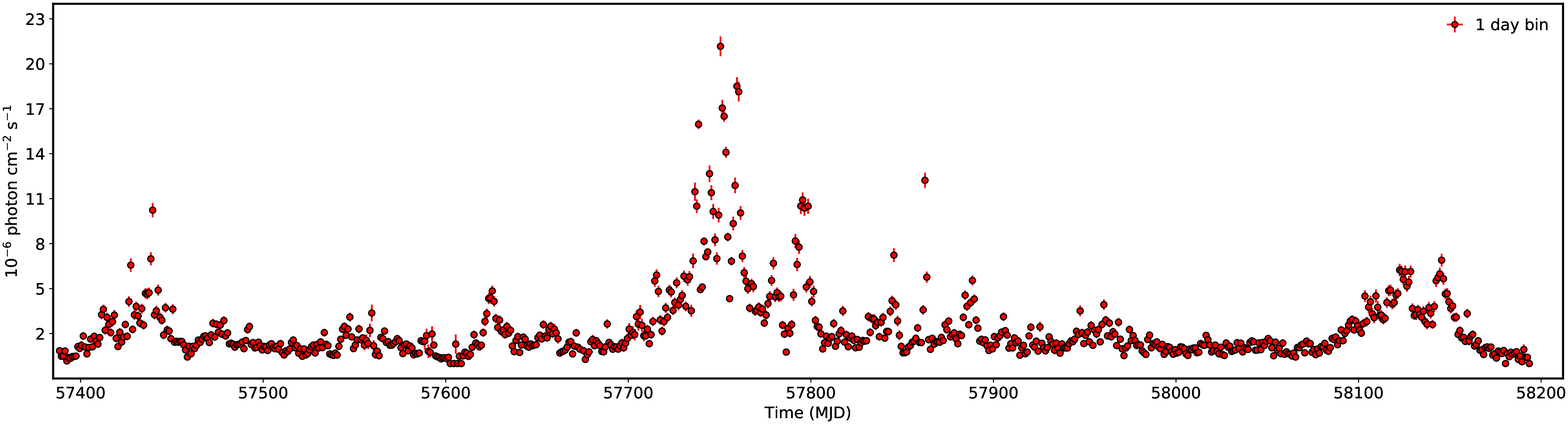}\\
   \includegraphics[width=0.48 \textwidth]{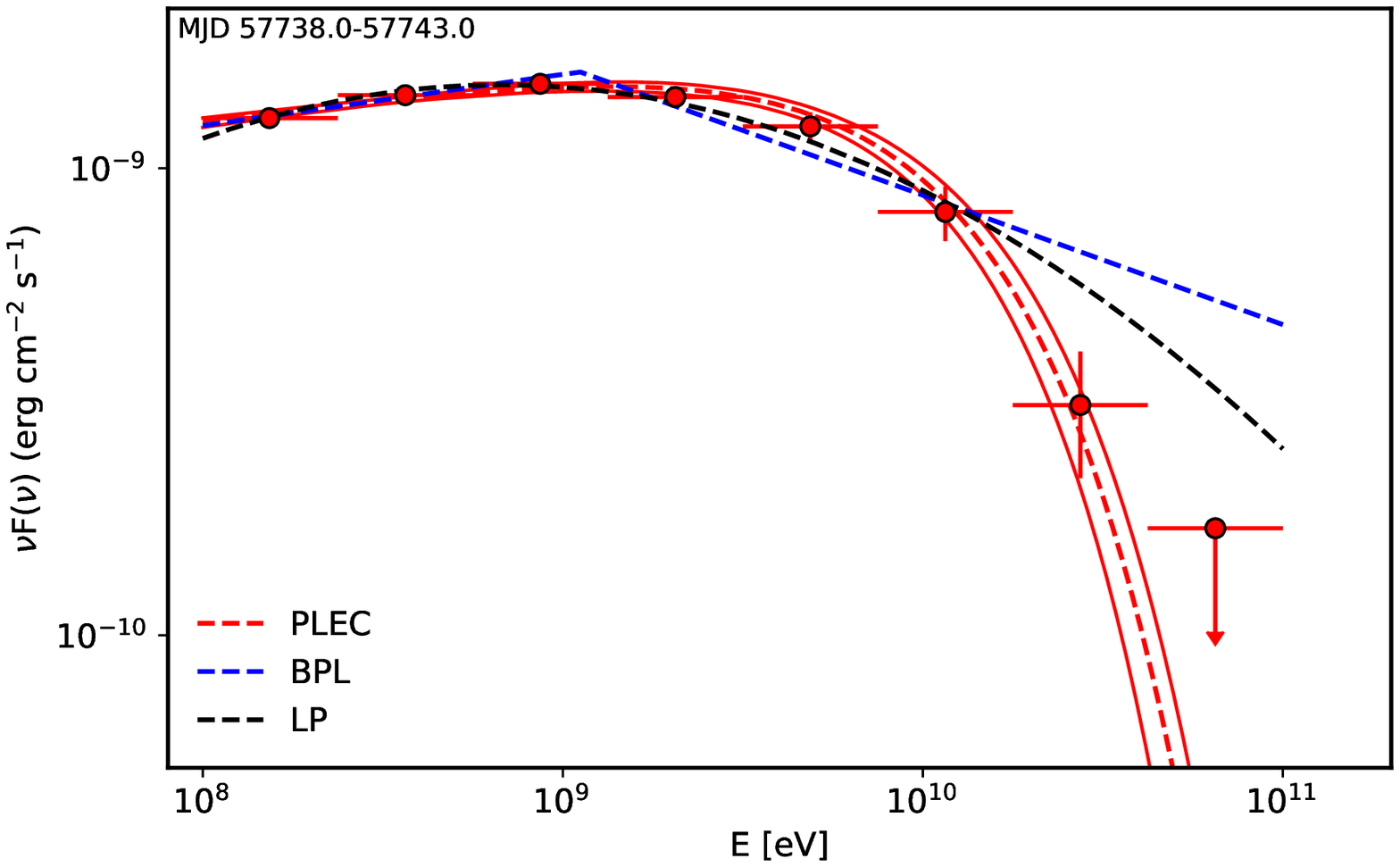}
      \includegraphics[width=0.48 \textwidth]{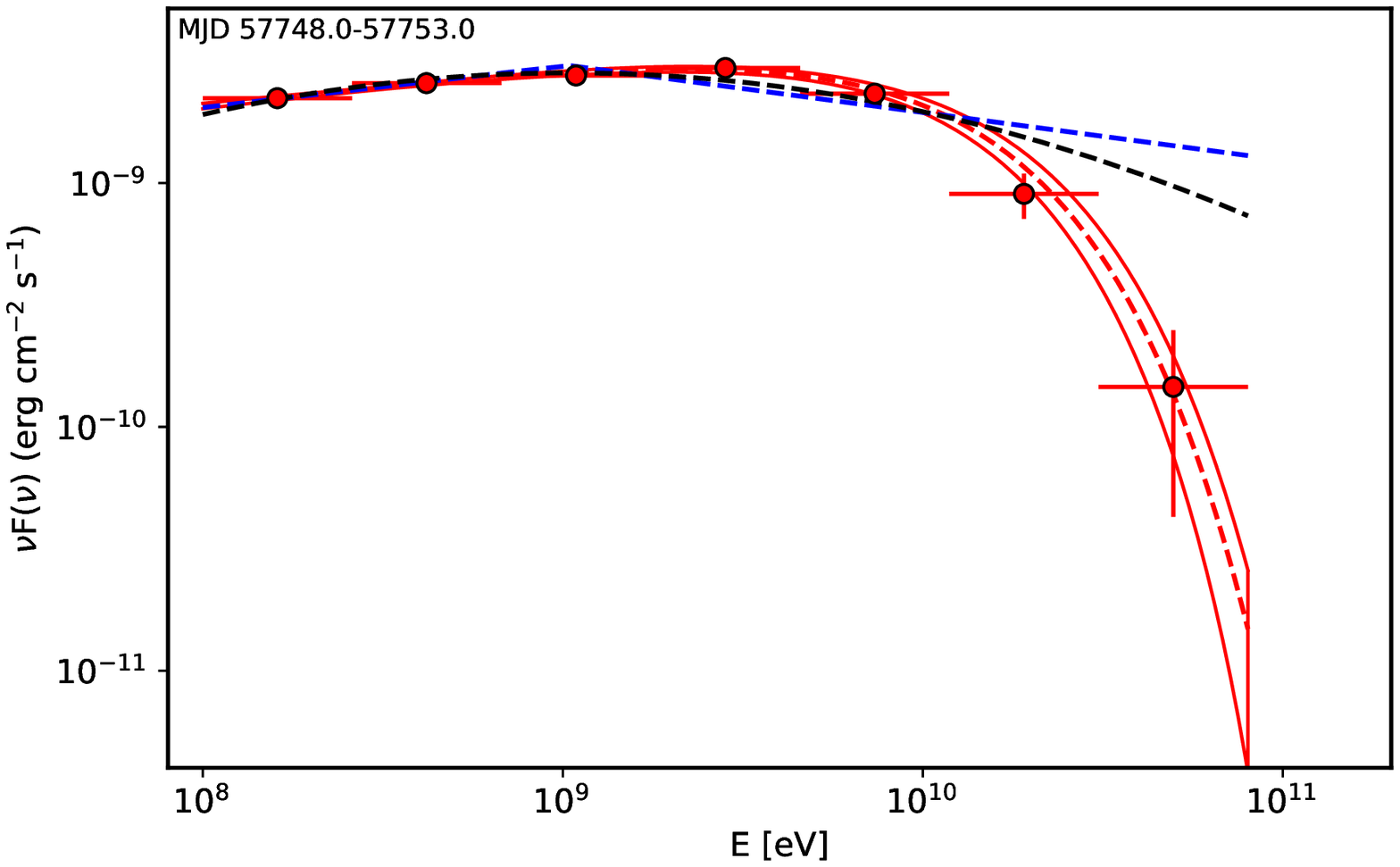}\\
      \includegraphics[width=0.48 \textwidth]{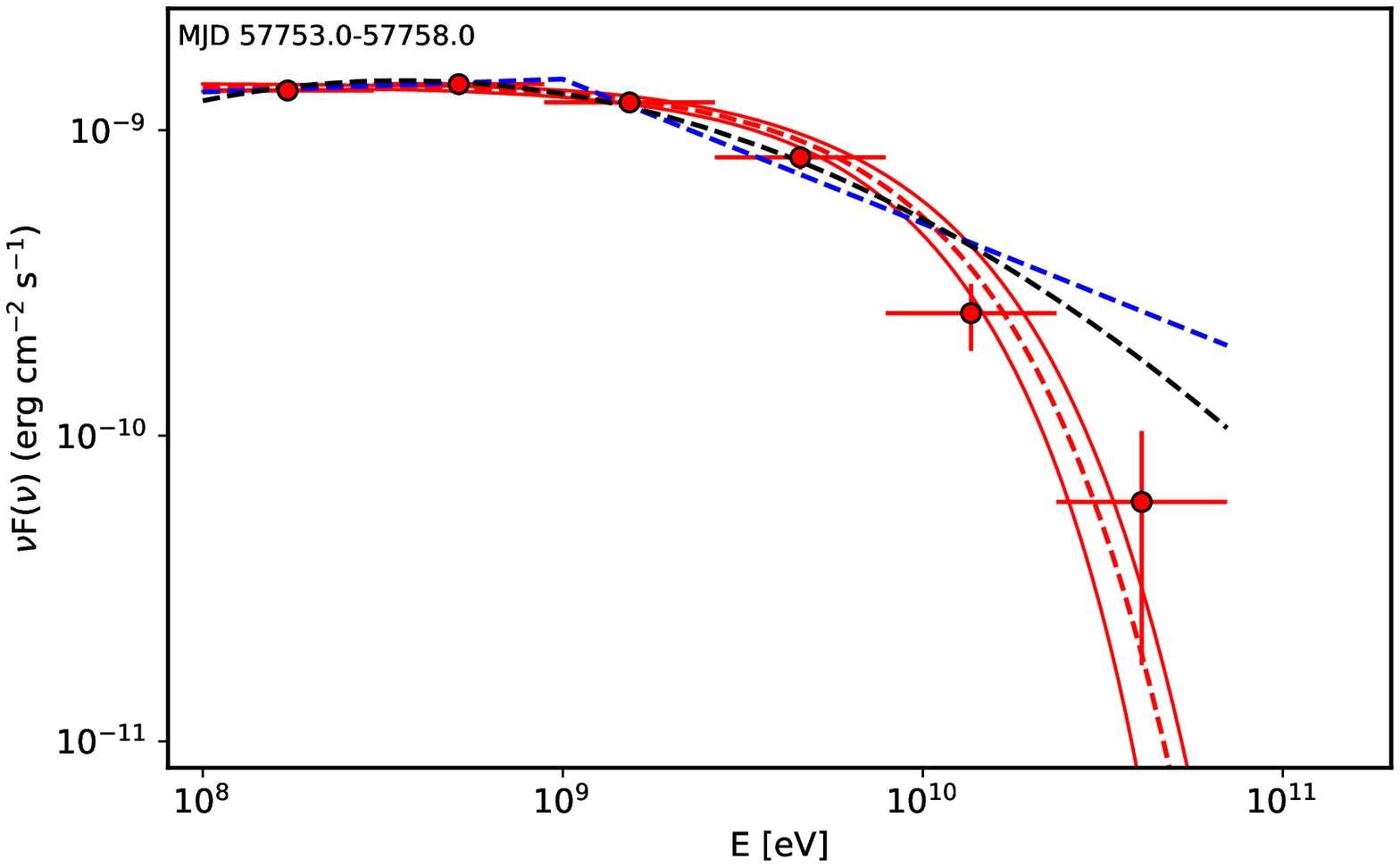}
      \includegraphics[width=0.48 \textwidth]{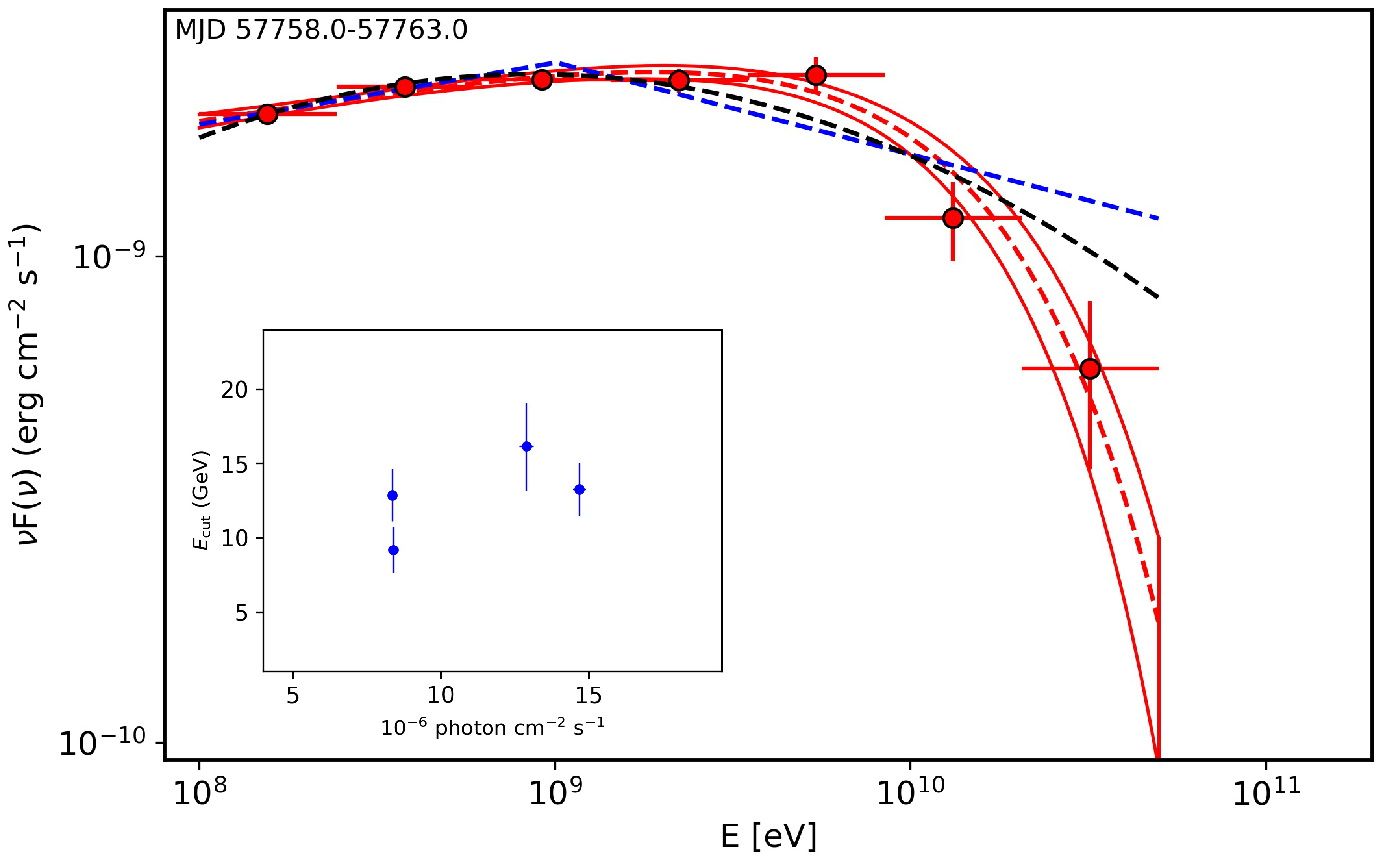}
   \caption{{\it Upper Panel:} The \gray light curve of \ct with 1-day time bins from 01 January 2016 to 01 April 2018. {\it Lower panels:} The \gray spectra in the energy range from 100 MeV to 300 GeV for the periods which showed significant deviation from the simple power-law model. The power-law with exponential cut-off spectral model (dashed red line) with the fit uncertainties (red solid lines) are shown together with the spectral points and is compared with other adopted models (broken power-law in blue and log parabola in black). The spectral points are obtained by separately running gtlike for smaller energy intervals.}
    \label{fg1}
\end{figure*}
\section{Fermi LAT observations}\label{sec2}
The Fermi LAT data accumulated from 01 January 2016 to 01 April 2018, when the large-amplitude flaring activities of CTA 102 occurred, are analyzed. LAT on board the Fermi satellite is a pair-conversion telescope sensitive to HE \grays in the 20 MeV - 300 GeV energy range \citep{atwood}. Collecting the data since 2008,  it is scanning the entire sky every $\sim$3 hours, thereby providing most detailed view of nonthermal HE processes occurring in the astrophysical sources. The {\it PASS8} version of the data in the energy range between 100 MeV - 300 GeV were analyzed using Fermi LAT Science Tool version 1.0.10 with the instrument response function P8R2\_SOURCE\_V6. The entire data set is filtered with {\it gtselect} and {\it gtmktime} tools and only the events with a high probability of being photons {\it evclass=128, evtype=3} have been considered. The zenith angle cutoff $>90^{\circ}$ is chosen to exclude atmospheric \grays from the Earth limb that can be a significant source of background. The data downloaded from a region defined as a circle of a $12^{\circ}$ radius centered at the \gray position of \ct (RA, Dec) = (338.152, 11.731) are binned within a $16.9^{\circ}\times16.9^{\circ}$ square region with {\it gtbin} tool with a stereographic projection into $0.1^{\circ}\times0.1^{\circ}$ pixels. The model file describing the region of interest is generated using the \fermi{} fourth source catalog \citep{2019arXiv190210045T} (4FGL) where the sources within $12^{\circ}+5^{\circ}$ from the position of CTA 102 are included. The model file contains also the standard Galactic {\it gll\_ iem \_ v07} and isotropic {\it iso\_P8R3\_SOURCE\_V2\_v1} diffuse components. The normalization of background models as well as the fluxes and spectral indices of the sources within $12^{\circ}$ are left as free parameters during the analysis.
\begin{table*}
\caption{The parameters of fitting with PLEC model in the periods showing deviation from the power-law model.}
\centering
\begin{tabular}{l c c c c c} \hline
Period & $\Gamma$  & ${\rm E_{\rm cut}}$&$F_{100}$ & $\sqrt{2(\Delta\mathcal{L})}$\\
(MJD )   & & (GeV)& (10$^{-6}$ ph cm$^{-2}$ s$^{-1}$)& &  \\\hline
57738-57743 & $1.89\pm0.02$  & $12.86\pm1.77$ & $8.37\pm0.12$ &10.41\\
57748-57753 & $1.84\pm0.01$  & $13.24\pm1.79$ & $14.69\pm0.22$ &10.66\\
57753-57758 & $1.98\pm0.02$  & $9.18\pm1.54$ & $8.39\pm0.14$ &8.30\\
57758-57763 & $1.88\pm0.02$  & $16.12\pm2.98$ & $12.89\pm0.24$ &7.65\\
\hline
\end{tabular}
\label{tab_SD}
\end{table*}
Initially, the binned likelihood analyses is applied to the full time data set adopting a log-parabola spectrum for \ct, however, for the light curve calculations a power-law model was used. The photon indices of all background sources were fixed to the obtained best guess values in order to reduce the uncertainties in the flux estimations in short periods. The \gray light curve is calculated with the unbinned likelihood analysis method implemented in the {\it gtlike} tool with the appropriate quality cuts as applied in the data selection. The \gray{} light curve with one-day binning is shown in Fig. \ref{fg1}. An interesting evolution of the \gray flux can be noticed: the source is in its flaring state alternatingly, with the highest flux being observed on 57750 MJD which corresponds to $(2.12\pm0.07)\times 10^{-5}{\rm photon\:cm^{-2}\:s^{-1}}$ while the hardest photon index is $1.80\pm 0.06$ observed on MJD 57424. This source is variable in timescales less than a day, however, for shorter periods the observed spectra will extend only up to moderate energies, preventing detailed spectral analyses. Since here the curvature in the \gray spectra of \ct is intended to study, the periods $>1$ day are considered to gather sufficient statistics. For a detailed study of the \gray light curve of \ct in short and long time scales as well as in the multiwavelength context see \citet{2018ApJ...863..114G}.\\
The spectra of \ct in 0.1-300 GeV band are investigated by detailed spectral analyses. In order to identify the periods where the spectrum significantly deviates from a simple power-law model, the data with different time binning (from 1 to 6 days) were analyzed. Yet, the time-averaged \gray spectrum of \ct is characterized by a soft photon index with a smooth break at higher energies. So, there were further considered only the periods when a harder photon index was observed (e.g., $\Gamma\leq$2.1). This allows to select from the flaring states only the periods exhibiting substantially different properties as compared with those observed in the quiescent state. Then, for each period, plots of Counts/bin versus Energy and residuals between the model and the data are generated, comparing the assumed power low spectrum with the observed data. Among the selected periods, when the power-law model reasonably well explains the observed data have been excluded, so there remain only the periods where a hint of a possible deviation from a power-law model is present.  Then, in order to check for a statistically significant curvature in the spectrum, an alternative fit with the following functions were considered:\\
a power law with an exponential cut-off (PLEC) in the form of
 \begin{equation} \label{PLEC}
  dN(E)/dE = N_0 (E/E_0)^{-\Gamma} \exp(-E/E_c),
  \end{equation}
a log-parabola (LP), defined as
 \begin{equation} \label{LP}
 dN(E)/dE = N_0 (E/E_0)^{-\alpha-\beta\ln(E/E_0)},
 \end{equation}
and a broken power law (BPL), defined as
 \begin{equation} \label{4}
  dN(E)/dE = \begin{cases} (E/E_{\rm b})^{\Gamma_1}, & \mbox{if } E < E_{b} \\ (E/E_{\rm b})^{\Gamma_2}, & \mbox{if } E > E_{b} \end{cases}
  \end{equation}
Different models are compared using a log likelihood ratio test, when the significance is estimated as twice the difference in the log-likelihoods. The spectral parameters of \ct are considered as free parameters during the analyses while the photon indices of all sources within the ROI are fixed to the values obtained during the whole analysis. The best matches between the spectral models and events are obtained using an unbinned analysis method. Then, the spectrum of \ct for each period was calculated by separately running {\it gtlike} tool for equal logarithmically-spaced energy bins.\\
The spectral models given in Eq. \ref{PLEC}-\ref{LP} are used to model the spectrum of \ct in each single period, and the significance of the curvature was estimated by comparing each model with the power-law. Although, almost in all the considered time intervals (from one to six days) a statistically significant curvature in the \gray spectra was observed, the most significant it is in five-day bins. The \ct spectra deviating from a simple power-law model with a significance exceeding $5\sigma$ are shown in the lower panel of Fig. \ref{fg1} and the corresponding parameters are given in Table \ref{tab_SD}. The data fitted with PLEC (red), BPL (blue) and LP (black) models are shown. As the fitting provides only log-likelihood values the models cannot be directly compared, so the goodness of the fit ($\chi^2$), which compares the data points with the models, is computed. This shows that the PLEC model is preferable for all the periods; other models yield a noticeably worse fit. These periods are characterized by a relatively hard photon index ($\Gamma=1.84-1.98$) and a cut-off around tens of GeV which does not change significantly in different periods ($E_{\rm cut}=9.40-16.12$ GeV). The variation of $E_{\rm cut}$ with the flux is shown in the inset of the lower panel in Fig. \ref{fg1}. In the considered period the flux and cut-off are not varying significantly. Similar conclusion can be drawn when BPL model is considered (although it fails to explain the data observed at higher energies): the break energy varies around $E_{\rm br}\simeq1.0$ GeV. 
\begin{figure*}
  \includegraphics[width=0.48 \textwidth]{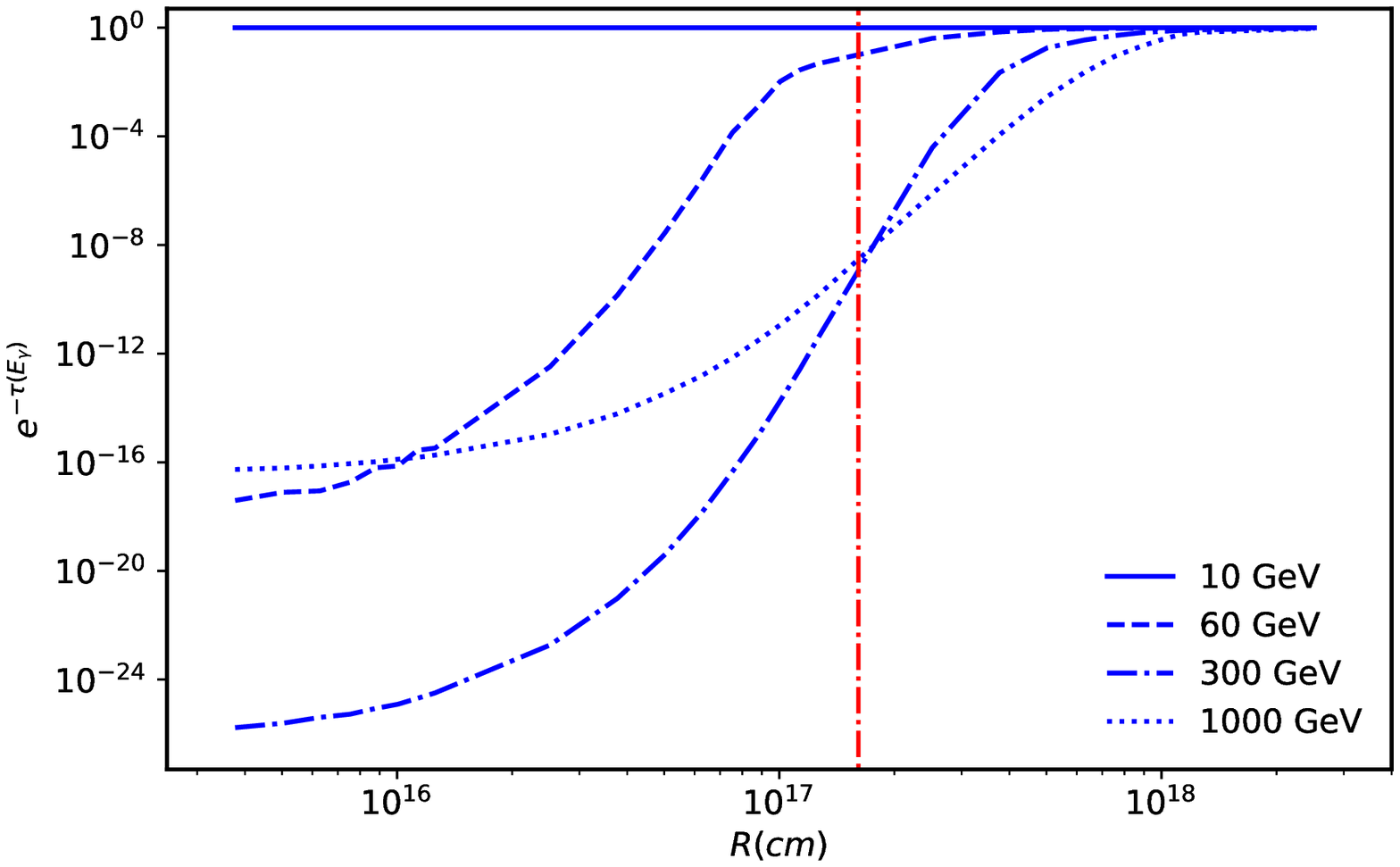}   
   \includegraphics[width=0.48 \textwidth]{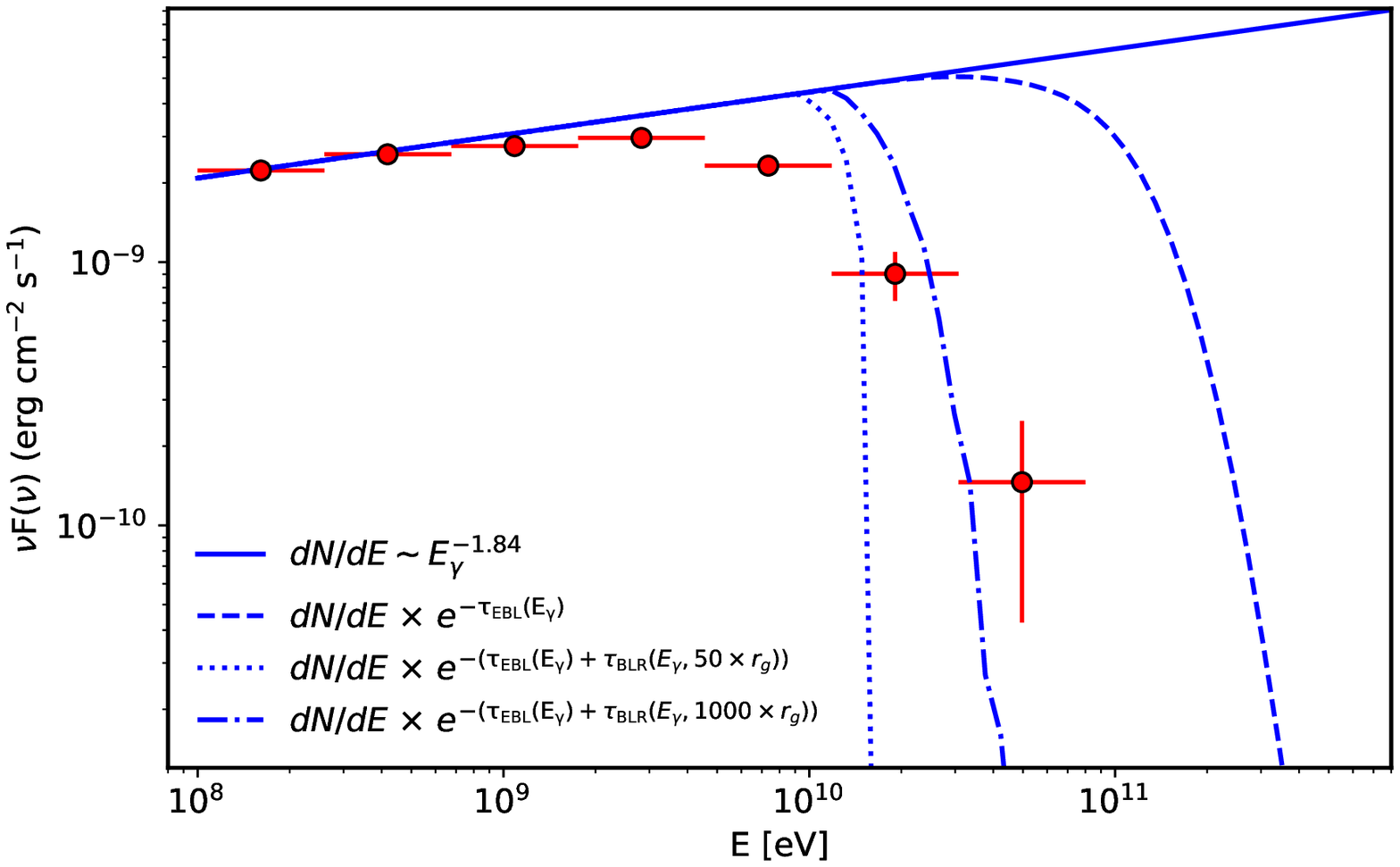}    
   \caption{{\it Left panel:} Internal BLR absorption as a function of distance for different \gray energies. The red dot-dashed line shows the $R_ {Ly\alpha}$ radius. {\it Right panel:} The reconstructed power-law model compared with the data considering external (EBL) and internal absorptions. The latter is computed assuming the emission region is at $\sim50\:r_{\rm g}$ (doted blue line) and at $\sim1000\:r_{\rm g}$ (dot-dashed blue line) distances from the central source.}
    \label{fg2}
\end{figure*}
\section{Absorption of \grays}\label{abs}
The curvature of the \gray spectrum of \ct, reported in the previous section, can be of different origin. In principle, it can be due to absorption, when the GeV \grays interact with the low energy photons (through $\gamma\gamma$ collision) or can be related with similar steepening in the spectrum of the emitting particle distribution due to the interplay of acceleration and cooling processes. Understanding the exact nature of this steepening can help to investigate the processes taking place in the jet of \ct or can help to localize the \gray emitting region.\\
The \grays  can be absorbed either inside the source interacting with the photons reprocessed from BLR or during their propagation interacting with extragalactic background light (EBL) photons. Considering, the distance of of \ct (z=1.037), the absorption due to interaction with EBL photons is significant for energies $\geq (200-300)$ GeV as shown in in the right panel of Fig. \ref{fg2} (dot-dashed blue line) where the extrapolation of only the power-law component ($\sim E^{-1.84}$) observed in MJD 57748-57753 is corrected for EBL absorption using the model from \citet{2011MNRAS.410.2556D}. Such absorption cannot explain the observed steeping of the spectrum at lower energies. In addition, if the emitting region is inside the BLR, the photons can be also effectively absorbed when interacting with the optical photons. Following the treatment of \citet{2016ApJ...830...94F}, the optical depth is calculated by modeling the BLR as infinitesimally thin spherical shells or thin rings. The luminosity and radius of the shells or rings are estimated using the composite quasar spectrum from the SDSS \citep{2001AJ....122..549V} in terms of ${\rm L_{H\beta}}$ luminosity (which is ${\rm L_{H\beta}}=(8.93\pm6.00)\times10^{43}\:{\rm erg\:s^{-1}}$ for \ct \citep{2019ApJ...877...39M}). The absorption is dominated by ${\rm Ly\alpha}$ photons at the radii $1.61\times10^{17}$ cm (see \citet{2016ApJ...830...94F}, for further details) although the absorption by other lines is not negligible. The absorption by the photons directly from the accretion disk is not considered as it is significant at $\geq{\rm TeV}$ \gray energies \citep{2003APh....18..377D, 2007ApJ...665.1023R, 2016ApJ...830...94F}. The attenuation ($e^{-\tau(E_{\gamma}, R)}$) strongly depends on the distance from the central object and energy of photons. For example, the plot of attenuation versus the distance is shown in Fig. \ref{fg2} (left panel) for different distances of emitting region and for photons with energies 10, 60, 300 and 1000 GeV. The region is optically thin for 10 GeV photons (blue line in Fig. \ref{fg2}) which escape it unabsorbed. Instead, the higher-energy photons will be heavily absorbed when the emitting region is inside the BLR (dashed, dot-dashed and dotted blue lines in Fig. \ref{fg2} left panel). The absorption decreases at larger distances making a small contribution at $>10 R_ {Ly\alpha}$. Note that similar result was obtained for a different geometry of BLR \citep{2019ApJ...871...19Z}.\\
The effect of attenuation due to the interaction with BLR photons in the extrapolated power-law spectrum for different distances of the emitting region is shown in Fig. \ref{fg2} (right panel) where a factor of $(1+z)$ is taken into account for the energy as the absorption is in the galaxy frame. When the compact emitting region is at a distance of $R=50*r_{\rm g}$ (where $r_{\rm g}=1.26\times10^{14}{\rm cm}$ is the gravitational radius for the \ct black hole mass of $M=8.51\times10^8M_{\odot}$ \citep{2014Natur.510..126Z}), the emitted flux will sharply decrease at energies $> 10$ GeV and cannot explain the observed data (dotted line in Fig. \ref{fg2} right panel). When the region is close to the distance of BLR, $R=1000*r_{\rm g}$, the flux drops slowly but still can not describe the observed spectra (dot-dashed blue line in Fig. \ref{fg2} right panel): the model overproduces the flux observed around 10 GeV. For further distances, the absorption becomes less significant and the observed steepening cannot be interpreted by BLR absorption. On the other hand, the observed variability time-scales put an additional constraint on the distance of the emitting region. For example, in \citet{2018ApJ...866...16P} using a 6-hour binned light curve of Fermi LAT data, the flux doubling time around MJD 57752 is $5.05\pm0.85$ hours  \citep[flare 3 in][]{2018ApJ...866...16P} which implies that the size of the emitting region is constrained by $R_{\rm \gamma}\leq \delta \times c\times t_{\rm var}/(1+z)\simeq2.68\times10^{14}\times \delta\; {\rm cm}$ where $\delta$ is the Doppler factor which is equal to bulk Lorentz factor for small viewing angle ($\delta\simeq\Gamma$). Using VLBA data a bulk Lorentz factor of $\geq17.5$ and a jet half opening angle of $\theta_{\rm j}\leq1^{\circ}.8$ were estimated for \ct \citep{2018ApJ...854...17L}. So, if the entire jet width is responsible for the emission, the emitting region along the jet should be at the distance of $\sim R_{\gamma}/\theta_{\rm j}\simeq1.49\times10^{17}\: {\rm cm}$ which is close to the upper edge of BLR. As discussed above, at these distances the absorption is relatively weak and cannot account for the observed steepening.
\begin{figure*}
  \includegraphics[width=0.48 \textwidth]{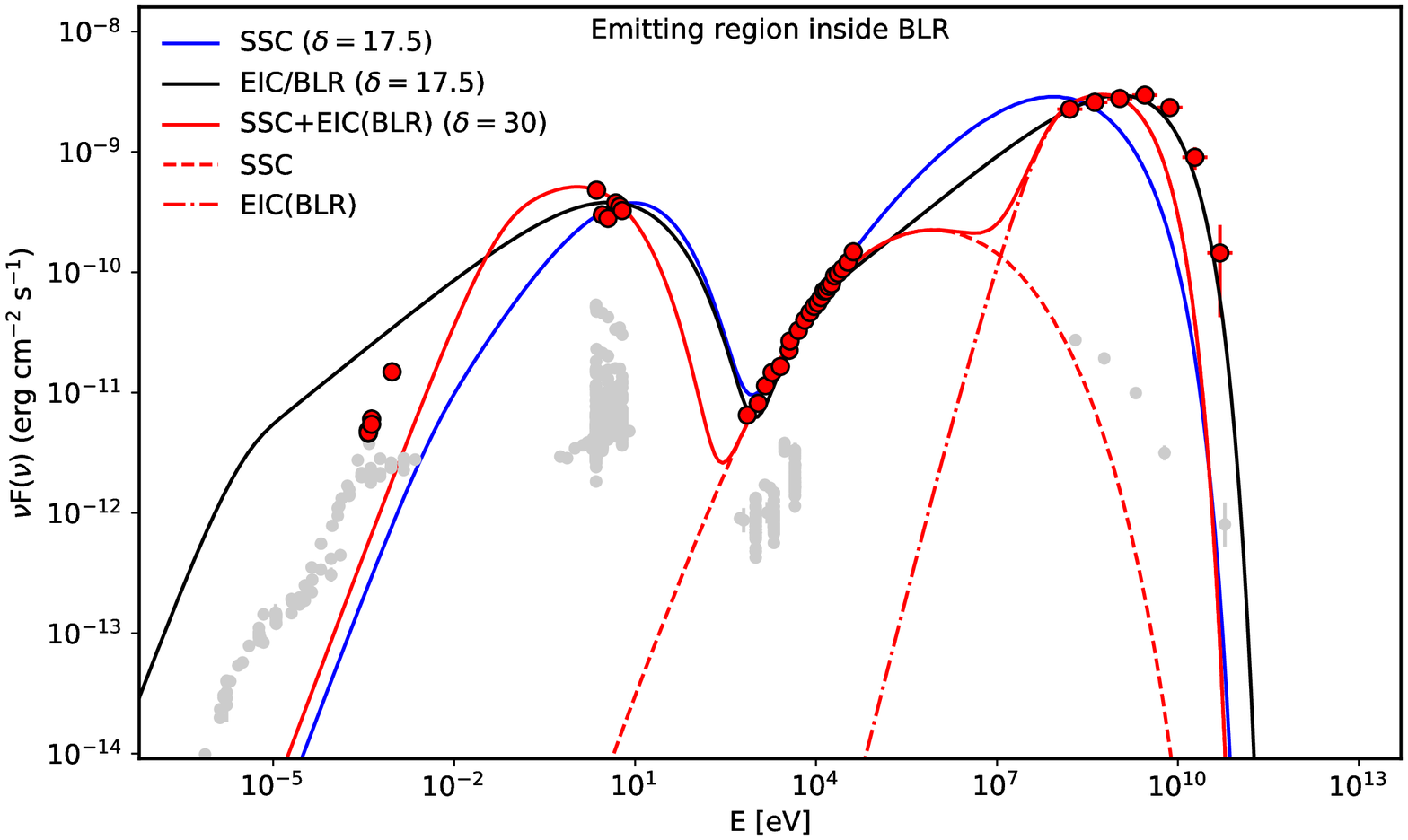}
  \includegraphics[width=0.48 \textwidth]{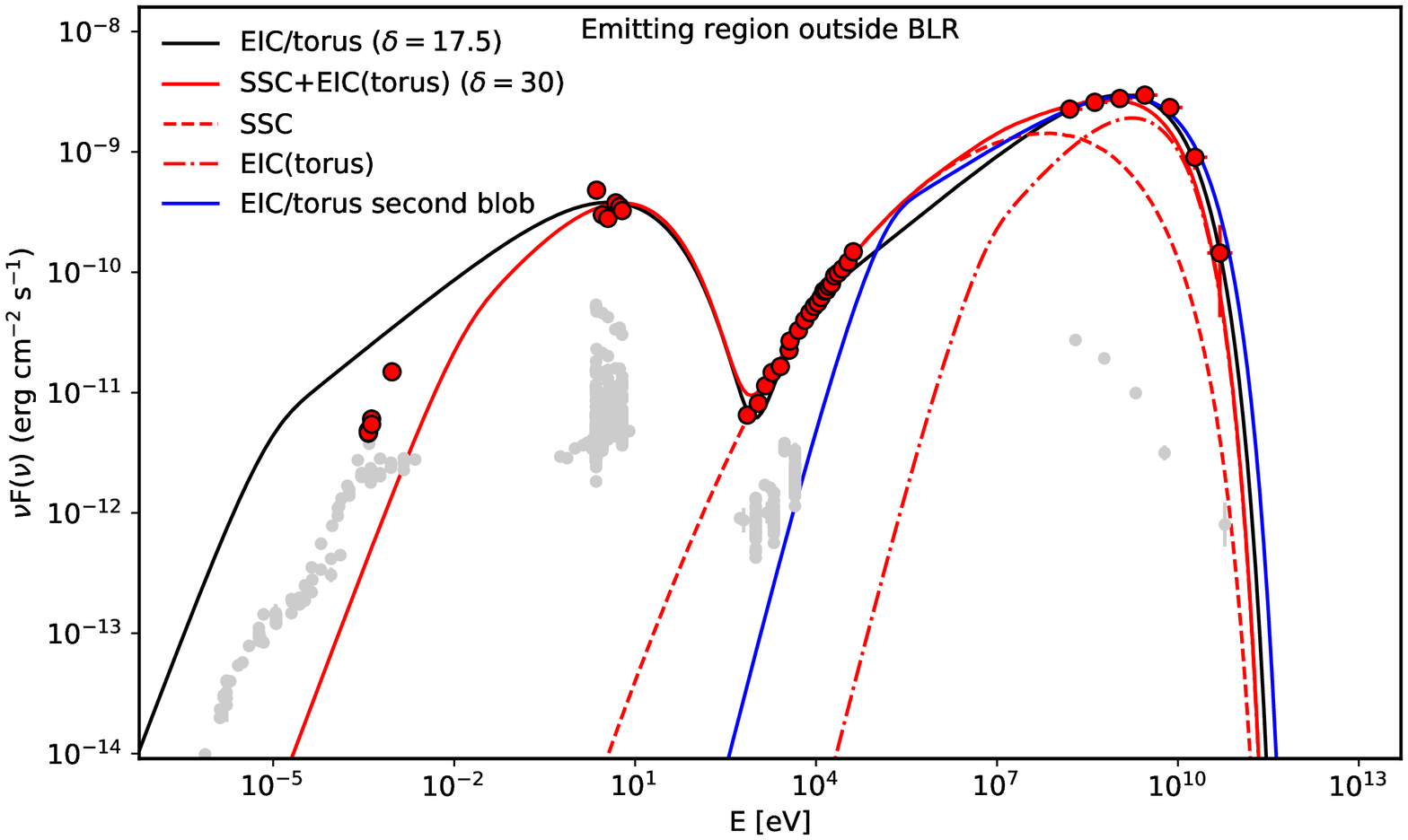}
   \caption{The SED of \ct for the period from MJD 57748 to MJD 57753. Th archival data are shown in light gray. The models are shown assuming the emitting region is inside the BLR ({\it left panel}) and outside the BLR ({\it right panel}). Different models are noted in the plot legend and the model parameters are given in the text.}
    \label{fg22}
\end{figure*}
\section{Origin of multiwavelength emission}\label{sec4}
Even at very close distances from the central source, the absorption due to the interaction with BLR photons cannot explain the observed steepening in the \ct spectra. These breaks are most likely related with an intrinsic break in the spectrum of the radiating particles (electrons). Considering the available large amount of multiwavelength data from the observations of \ct, their modeling is crucial for estimating the underlying particle energy distribution which in its turn allows to study the particle acceleration and cooling processes.\\
The multiwavelngth SED of \ct is shown in Fig. \ref{fg22} where the Swift UVOT/XRT data are from \citet{2018ApJ...863..114G} for the period 3 (MJD 57752.52). For the NuStar data, the same analysis described in \citet{2018ApJ...863..114G} was repeated but only the counts up to $45$ keV, where the X-ray spectra of \ct are above the background, were extracted. The \gray data are accumulated for the period from MJD 57748 to MJD 57753. The data in the mm/sub-mm band are from the ALMA catalog from the observations of bright compact radio sources in different bands between May 2011 and July 2018 \citep{2019MNRAS.485.1188B}. From the many observations of \ct, only the data from the observations carried out on December 17, 18 and 29, 2016 and on January 8, 2017, which are nearly simultaneous with the studied flare, were considered. For comparison, the time averaged data from \ct observations are shown in light gray, which highlights the changes observed in different energy bands.\\
A deviation from the power-law model has been observed in the spectra of several blazars \citep[e.g.,][]{2009ApJ...699..817A, 2010ApJ...710.1271A,2012ApJ...761....2H,2015MNRAS.449.2901K,2013A&A...557A..71R,2015ApJ...808L..48P,2015ApJ...803...15P, 2019ApJ...881..125D}. The internal or external attenuation cannot be responsible for the observed steepening as demonstrated in Section \ref{abs}. This is in agreement with the study of the spectra of 106 broad-line blazars detected in the MeV/GeV bands showing no evidence of expected BLR absorption \citep{2018MNRAS.477.4749C}. So, the steepening might be most likely caused either by a similar feature present in the spectra of the emitting electrons or by the transition of IC scattering from Thomson to Klein-Nishina regimes. The IC scattering occurs in the Klein-Nishina regime when $E_{\rm e}^{\prime}> (m_{\rm e}\:c^{2})^2/4/3 \epsilon^\prime_{0}$. So, when the IC scattering of synchrotron photons (peaking in infrared to optical bands) or IR photons from dusty torus is considered, the IC scattering to MeV/GeV energies typically occurs in the Thomson regime. In contrast, when BLR photons are considered, the IC scattering to the same energies is in the Klein-Nishina regime. The break energy in the \gray spectrum naturally formed by the Klein-Nishina effects on the Compton scattering depends on the target photon energy and is independent of $\delta$ \citep{ghiselini09, 2010ApJ...721.1383A}. By considering different values for the target photon field, the cut-off at energies observed for \ct can be reproduced by the Klein-Nishina effect. However, considering the limited information available on the BLR photons, this would be based only on inferred assumptions rather than on a real physical picture. Instead, if the break is caused by the particles and when the parameters describing the energy distribution of the particles are constrained, the physics of jets can be explored. To keep generality, during the modeling different distances for the emission region is assumed (inside and outside BLR) and all the relevant photon fields as well as the Klein-Nishina effects on the IC scattering are taken into account.\\
A one-zone leptonic emission scenario was used assuming that the emitting electrons are confined in a compact spherical region with a radius of $\simeq2.68\times10^{14}\times\delta\:{\rm cm}$ and magnetic field intensity of $B$. Due to relativistic motion of the jet, the radiation will be Doppler boosted by $\delta=\Gamma\geq17.5$ \citep{2018ApJ...854...17L} and will appear brighter for the observer. For the underlying particles a PLEC distribution within $E_{\rm min}^\prime$ and $E_{\rm max}^\prime$ is assumed:
\begin{equation}
N(E_{\rm e})= (E_{\rm e}/m_e c^2)^{-\alpha}exp(-E_e/E_c)\: [{\rm eV^{-1}}]
\label{EPLC}
\end{equation}
considering the the total energy of electrons, $U_{\rm e}=\int_{\rm E_{min}^\prime}^{\rm E_{\rm max}^\prime}E_{e}^\prime N_e(E_{e}^\prime)dE_e^\prime$, as a free parameter during the fitting. The free model parameters are estimated using Markov Chain Monte Carlo (MCMC) method which enables to derive the confidence intervals for each parameter (the application of the method and the used code are described in \citet{2017MNRAS.470.2861S} and \citet{2018ApJ...863..114G}).
\subsection{Emitting region inside the BLR}
When the emitting region is inside the BLR, the dominant photon fields which are IC up-scattering to X-ray-\gray bands are synchrotron photons and disc-emitted photons reflected from the BLR. The IC scattering of only synchrotron photons with $\sim1$ eV peak energy on the electron population with an energy distribution with a cut at $E_{\rm cut}\leq1.6\times (B/{\rm 1 G})^{-1/2}\times (\delta/17.5)^{-1/2}\: {\rm GeV}$ (constrained from $E_{\rm s, peak}\leq 1\: {\rm eV}$) will extend only up to $(1.8-2)$ GeV which is insufficient to explain the observed data (see blue line in Fig. \ref{fg22} left panel). Considering 10\% of the disc emission is reflected from BLR with a radius of $R_{\rm BLR}=10^{17}\:(L_{\rm d}/10^{45})^{0.5}=6.3\times10^{17}$ cm (where $L_{\rm disc}=10\times L_{\rm BLR}\simeq4.0\times10^{46}\:{\rm erg\:s^{-1}}$ \citep{pian}), the external photon field density in the jet frame will be $U_{\rm BLR}= L_{\rm BLR}\:\delta^2/4 \pi R_{\rm BLR}^2 c= 0.026\times \delta^2\:{\rm erg \: cm^{-3}}$. This will over-exceed the synchrotron photon density when high Doppler boosting is assumed, e.g., $\delta=30$ which is more typical for powerful blazars. In this case, when $\alpha=1.81\pm0.09$, $E^{\prime}_{\rm cut}=0.37\pm0.04$ above $E^{\prime}_{\rm min}=76.10\pm2.10$ MeV and the magnetic field in the emitting region is $B=8.24\pm0.18$ G, the EIC peaks around GeV energies, explaining the \gray data, while the X-rays are due to SSC radiation (red line in Fig. \ref{fg22} left panel). Because of the high magnetic field necessary to explain the UV and X-ray data by synchrotron/SSC processes, the electron distribution should have a lower cut-off energy ($E^{\prime}_{\rm cut}=0.37$ GeV) which does not allow satisfactory modeling of the observed data at HEs. In this case the jet is magnetic field dominated with $U_{\rm e}/U_{B}=0.06$. In principle the magnetic field can be reduced by increasing the total energy of the emitting electrons, in which case the IC will overproduce the \gray data below $\sim1$ GeV.\\
The required magnetic field can be decreased in an alternative model where the X-ray to \gray emission is due to IC up-scattering of only BLR photons (black line in Fig. \ref{fg22} left panel).
Then, when $\delta=17.5$, the estimated magnetic field is lower, $B=3.68\pm0.04$ G, and $E^{\prime}_{\rm cut}=2.02\pm0.04$ GeV with $\alpha=2.18\pm0.003$ allowing to model the observed data. The low-energy tail of the HE component can reproduce the X-ray data only at lower $E^{\prime}_{\rm min}=1.1\pm0.01$ MeV (normally it is expected that $\gamma_{min}=E_{\rm min}/m_e c^2$ should be close to unity \citep{cellot}). However, the synchrotron emission of such low-energy electrons will overproduce the observed radio flux, but one should note that synchrotron-self absorption is not taken into account, which is significant below $4\times 10^{-2}$ eV \citep{2018ApJ...863..114G}.
\subsection{Emitting region outside the BLR}
When the emitting region is beyond the BLR (e.g., at $> 0.2\:{\rm pc}$), the IR photons from the dusty torus ($R_{\rm IR}=10^{18}\:(L_{\rm d}/10^{45})^{0.5}=6.32\times10^{18}$ cm \citep{ghiselini09}) with $U_{\rm IR}= L_{\rm IR}/4 \pi R_{\rm IR}^2 c\:\delta^2= 1.59\times10^{-3}\times\delta^2\:{\rm erg \: cm^{-3}}$ density will dominate over that of BLR-reflected photons which will decrease as $\sim U_{\rm BLR}/(1+(R/R_{\rm BLR})^3)$ beyond $R_{\rm BLR}$. In Fig. \ref{fg22} right panel, the SED modeling when both the synchrotron and torus photons are considered is shown with red solid line (for $\delta=30$). Again, as in the previous case, a high magnetic field $B=2.39\pm0.04$ G is required (although slightly lower as the energy density of torus photons compared with BLR photons is lower), and the X-ray data can be explained by SSC mechanism. As the average energy of IR photons with $\sim10^3$ K temperature is lower than that of BLR photons ($\sim10^{4}$ K), their IC up-scattering ($\sim \delta \gamma^2 (k_{\rm b}\:T)$) can explain the observed \gray data when $E^{\prime}_{\rm cut}=1.51\pm0.17$ GeV; the synchrotron emission of these electrons will slightly overproduce the soft X-ray data (red solid line in the right panel of Fig. \ref{fg22}).\\
A fit, assuming the X-ray and \gray data are due to IC up scattering of only torus photons, is shown in the right panel of Fig. \ref{fg22} (black solid line). As the magnetic field is low, $B=1.13\pm0.01$ G, the SSC component falls below the observed X-ray data. The power-law index of underlying electrons, $\alpha=2.18\pm0.004$, is defined by joint X-ray and \gray data (see next section). The minimum and cut-off energies of underlying electrons are estimated to be $E_{\rm min}^{\prime}=3.48\pm0.04$ MeV and $E_{\rm cut}^{\prime}=3.60\pm0.09$ GeV, respectively.\\
Further, it is assumed that the radio-optical-X-ray and \gray emissions are produced in different regions  (blobs). This permits  to estimate the properties of emitting electrons based only on the \gray data, without considering the effect of the magnetic field. Such consideration is motivate by the following: {\it i)} the previous studies of this source showed that the regions outside the torus are more favorable for the \gray emission \citep{2018ApJ...863..114G} and {\it ii)} the two-zone models were successful in explaining the bright flares of FSRQs \citep{2011A&A...534A..86T}. In this case, the power-law index of the emitting electrons is $\alpha=2.36\pm0.07$, much softer than in the previous cases which results in a larger cut-off energy $E_{\rm cut}^{\prime}=5.32\pm0.75$ GeV. Since the X-ray data are considered as an upper limit, a larger $E_{\rm min}^{\prime}=18.52\pm8.43$ MeV is obtained (red dot-dashed line in the right panel of Fig. \ref{fg22}). The jet should be very strongly particle dominated, $U_{\rm e}/U_{B}>>1$ and the synchrotron emission of these electrons in the second region will not make a significant contribution to the low energy band. Such a modeling gives independent information on the particle content and distribution, as the luminosity of IC scattering depends only on $N_{\rm e}$ as distinct from the synchrotron or SSC components when the luminosity depends on the product of $B^2$ and $N_{\rm e}$. Therefore, this provides straightforward information on the jet-accelerated particles.\\
The results obtained above do not significantly differ from those obtained in the previous studies of \ct within one-zone leptonic scenarios \citep[e.g.,][]{2018ApJ...866...16P, 2018ApJ...863..114G}. However, it is impossible to compare the obtained parameters directly because different values of emitting region size, Doppler boosting, etc. were used in the mentioned studies. The change in the initial set of the model parameters impacts the estimation of other free parameters of the model. Also, the SED considered and modeled here differs from the previously modeled ones and for different periods the model parameters might vary.
\section{Particle Acceleration and Energy losses}\label{sec5}
One of the effective ways to study the physics of the jets is through the modeling of their multiwavelength emission spectra. The applied models can reproduce/explain the data observed in a short time period which is not enough for understanding the global processes occurring in jets given the extremely variable character of their emission. However, any model trying to reproduce the transfer of the radiative output along the jet propagation should be able to explain single snapshots of the SEDs. The SED discussed here is important as the curvature in the \gray spectrum is most likely related with the similar feature in the emitting electron distribution, giving a chance to explore the particle acceleration and cooling mechanisms.\\
\begin{figure}
  \includegraphics[width=0.48 \textwidth]{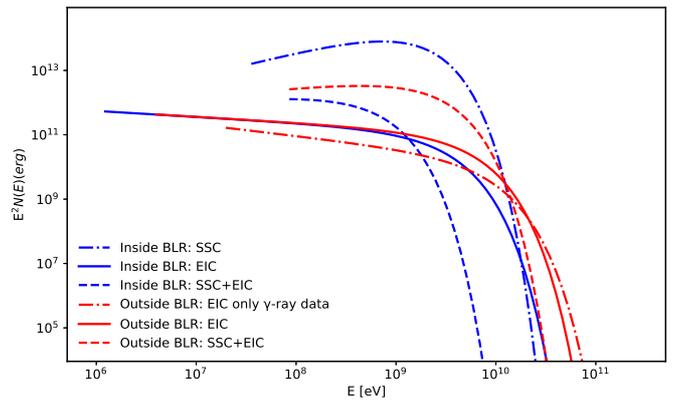}
   \caption{The energy distribution of the electrons for the models presented in Fig. \ref{fg22} obtained by MCMC modeling of the observed data.}
    \label{fg3}
\end{figure}
The electron energy distributions given by Eq. \ref{EPLC} that can explain the observed broadband emission of \ct are shown in Fig. \ref{fg3}. The free model parameters ($U_{e}$, $B$, $\alpha$ and $E^{\prime}_{\rm c}$) were extracted straightforwardly from the observed data using the MCMC method. In order to investigate the conditions for the formation of the electron energy distribution a detailed simulation of both acceleration of particles and treatment of temporal evolution of electrons taking into account relevant energy losses (e.g, solving the kinetic equation) are required \citep[e.g.,][]{1999MNRAS.306..551C}. Anyway, this is beyond the scope of this paper. Based on the estimated parameters, here an attempt is made  to put only quantitative constraints on the physical processes at work in the jet of \ct.\\
It is widely discussed that one of the most efficient mechanisms for energizing the particles in the relativistic jets of blazars is the first-order Fermi acceleration (diffuse shock acceleration [DSA]) \citep{1989MNRAS.239..995K, 1990ApJ...360..702E, 2012ApJ...745...63S}. A distinctive feature of this acceleration process (both for relativistic and non-relativistic shocks) is that the resulting particle energy distribution takes a power-law form ($E_{\rm e}^{-\alpha}$) \citep{1983RPPh...46..973D}. Under dominant radiative cooling and/or a decreasing chance for HE particles to cross the shock front a large number of times, the HE tail of the electron energy distribution steepens forming a power-law with an exponential cut-off distribution. So, the investigation of $\alpha$ and $E^{\prime}_{\rm c}$ parameters can shed a light on the physics of the jet.
\subsection{High energy cut-off in the electron spectrum}
The HE tail of electron distribution is defined by the cooling of emitting particles which in its turn strongly depends on the location of the emitting region. For example, when the emitting region is within the BLR having a higher density, the particle cooling is more efficient and they do not reach higher energies as compared to the case when the emitting region is outside the BLR (see Fig. \ref{fg3}). The cutoff electron energy is constrained by two conditions: {\it i)} the particles are not accelerated beyond the energies when the cooling and acceleration times are equal, and {\it ii)} the particles will not continue to accelerate beyond the energies permitted by the physical size of the emitting zone: $E_{\rm c}$ is determined as the smaller of these limiting values.\\
When the acceleration and cooling times are of the same order, the macroscopic parameters of the jet plasma start to play a crucial role and basically they are defining the formed spectrum of emitting electrons. In the collisionless shock the non-thermal particles are gaining energy by scattering between turbulence in the upstream and downstream of the plasma. 
The corresponding time for diffuse shock acceleration would be \citep{1983RPPh...46..973D, 2007Ap&SS.309..119R}
\begin{equation}
t_{\rm acc}\simeq \eta_0 \left(\frac{p}{p_{1}}\right)^{\alpha_{\rm diff}-1} \frac{m_{e} c \gamma_{\rm e}^{\prime}}{e B}\:\left(\frac{c}{u}\right)^2
\end{equation} 
where $p$ is the particle momentum, $\alpha_{\rm diff}$ is the diffusion index, $\eta\sim p^{\alpha_{\rm diff}-1}$ characterizes the diffusion ($\eta=1$ corresponds to Bohm limit), $u\sim c$ is the shock speed and $\gamma_{\rm e}^{\prime}=E^{\prime}_{\rm e}/m_{e}c^2$. Balancing this acceleration time with the electron cooling time defined as
\begin{equation}
t_{\rm cool}=\frac{3\: m_{e}c \:(1+z)}{4 \sigma_{\rm T}\:u_{\rm tot}^{\prime}\:\gamma_{\rm e}^{\prime}}
\end{equation} 
where $u_{\rm tot}^{\prime}=u_{\rm B}+u_{\rm SSC}+u_{\rm IR/BLR}$ and introducing cooling parameter $\epsilon_{\rm syn}$ defined as ratio of the luminosity of low energy component to the total luminosity, $L_{\rm low}/(L_{\rm low}+L_{\gamma})$, \citet{2017MNRAS.464.4875B} showed that the cutoff energy of the accelerated electrons scales with the magnetic field as $\gamma^{\prime}_{c}\simeq\sqrt{2 {\cal E}_s(\alpha_{\rm diff})/3 } \left(6\times 10^{15}/(\eta_0\, B) \right)^{1/(1 + \alpha_{\rm diff} )}$ where ${\cal E}_s(\alpha_{\rm diff} ) \simeq 3/2 \left( 9 \epsilon_{\rm syn}/4 \right)^{2/(1 + \alpha_{\rm diff} )} $ (assuming $u\sim c$). Through this equation $\alpha_{\rm diff}$ and $\eta_0$ are connected as $\eta_0\simeq 1.35\times10^{16}\, \epsilon_{\rm syn}\,B^{-1}\, \gamma_{\rm c}^{-(\alpha_{\rm diff}+1)}$ so when the magnetic field in the jet and the cutoff energy are known (e.g., from the multiwavelength data modeling) these parameters can be constrained. For \ct, when the emitting region is outside the BLR and the multiwavelength emission is described by synchrotron/SSC+EIC process then $B=2.39 \pm 0.04$ G and $E_{\rm c}^{\prime}=1.51 \pm 0.17$ GeV ($\gamma_c\simeq3.\times10^{3}$). For a fixed magnetic field and cutoff energy, $\eta_0$ scales inversely with $\alpha_{\rm diff}$, e.g., for Bohm type diffusion ($\alpha_{\rm diff}=1$) an unrealistically large $\eta_0\simeq10^{7}$ is needed (considering $\epsilon_{\rm syn}=0.14$). More relaxed  parameters are obtained when $\alpha_{\rm diff}>2$: $\eta_{0}\simeq 3\times10^4$ for $\alpha_{\rm diff}=2$ and $\eta_{0}\simeq 10$ for $\alpha_{\rm diff}=3.0$. Similarly, when the data are modeled by EIC of BLR photons, when $\alpha_{\rm diff}=2.0$ then $\eta_0$ is $8\times10^3$ and $\eta$ is $2.1$ when $\alpha_{\rm diff}=3$. These parameters indicate diffusion away from the Bohm limit with a stronger dependence of the mean free path on the momentum ($\sim p^{2}$). Larger value of $\eta_0$ implies that turbulence levels are gradually decreasing going farther from the shock. Similar values were obtained in the modeling of multiwavelength emission from BL Lacerte and AO 0235+164 when the DSA of particles was treated with detailed Monte Carlo simulations \citep[][]{2017MNRAS.464.4875B}, and $\eta_0=10^5$ was used to reproduce the broadband SED of Mrk 421 \citep{inou}.\\
When the dynamic time scales of the system are shorter than the acceleration times, the cutoff determined from $t_{\rm dyn}=t_{\rm cool}$ corresponds to $\gamma_{c}^{\prime}=3\: m_{e}c \:(1+z)/4 \sigma_{\rm T}\:u_{\rm tot}^{\prime}\:t_{\rm dyn}$. If the observed \grays are produced in a separate region under the dominant IC cooling of torus photons, the cutoff will be $\gamma_{c}^{\prime}=3\: \pi m_{e}c^2 \:(1+z) R_{\rm IR}^2/\sigma_{\rm T}\:\eta_{\rm IR}\:\delta^2 L_{\rm disc}\:t_{\rm dyn}$. For the variability time of the order of $5.05\pm0.85$ hours, the cut-off should be at $E_{e}^{\prime}=3.62$ GeV which is similar to the value estimated during the fit. This shows that the curvature in the electron spectrum might also come from a limitation from the acceleration zone.
\subsection{Power-law index of emitting electrons}
The power-law index of the emitting electrons is simulation-dependent and is strictly defined by the plasma parameters. Alternatively, it can be obtained through the modeling of the observed photon spectra, in some cases analytically as well. When the HE component is interpreted by IC up-scattering of the external photon field, the particle photon index is defined by $\alpha=2 \Gamma_{\gamma/X{\rm -ray}}-1$ \citep[e.g.,][]{2013LNP...873.....G}. In the case of \ct both X-ray and \gray data are defining the photon index ($\alpha_{\gamma/X{\rm -ray}}$) which can be obtained by fitting with a power-law function ($\sim(E/100\:{\rm eV})^{-\Gamma_{\gamma/X{\rm -ray}}}$). As for the BLR and torus photons IC scattering near the minimum energy of electrons ($\gamma_{\rm min}$ close to unity) is around $\sim (0.5- 0.7)$ keV and above $\sim 1$ GeV the \gray spectrum steepens, only the data observed between $\sim0.7$ keV and $\sim 1$ GeV are considered. The fit results in $\Gamma_{\gamma/X{\rm -ray}}=1.60\pm 0.01$, so $\alpha$ should be around $2.2$ which matches well with the estimated value of $2.18$. In the case of SSC+EIC scenario, the power-law index is mostly but not entirely defined by fitting the SSC component to X-ray data with a slope of $~1.32$. The SSC component can explain the X-ray data when assuming a hard $\sim1.6$ index for electron distribution but the EIC of these electrons will be steeper in the MeV/GeV band which is in disagreement with the observed data. The modeling resulted in a slightly different but still a hard spectrum for the electrons $\alpha=(1.7-1.8)$.\\
From the standpoint of shock acceleration theories the electron indexes discussed above can be easily formed under reasonable physical conditions. The DSA of particles establishes a power-law distribution of electrons with an index depending only on the shock velocity compression ratio ($\alpha=(r+2)/(r-1)$) \citep{1978MNRAS.182..147B, 1987PhR...154....1B, 1991SSRv...58..259J}. In the case of non-relativistic shocks with a large sonic Mach number $r=4$, so that the well known $E^{-2}$ spectrum will be formed. When relativistic shocks are considered the picture is changed because the assumptions made in deriving the spatial diffusion equation are no longer valid and the index is defined by the shock speed and also depends on the nature of particle scattering.  For a test particle in the parallel relativistic shocks the particles will be distributed by a universal power-law index of $-2.23$ \citep{2000ApJ...542..235K, 2004APh....22..323E, 1998PhRvL..80.3911B}. However, power-law indexes varying from very hard ($-1$) to very steep are possible, depending on the nature and magnitude of turbulence, shock speed and shock field obliquity \citep{2012ApJ...745...63S}. One of the best ways for studying the DSA of particles is through Monte Carlo simulations (although some analytical approaches were applied as well) making a detailed treatment of the shock speed, particle scattering, etc., which is beyond the scope of the current paper. The obtained power-law indexes from $1.8$ to $2.18$ are well within the values discussed for shock acceleration theories to date.
\section{Results and Discussion} \label{sec6}
The distinct blazar variability in almost all wavebands makes them ideal targets for exploring the particle acceleration and emission processes. Due to the processes causing the flares, the spectra of the sources sometimes exhibit dramatic changes in both amplitude and spectrum. So, the multiwavelength observations in these periods and their modeling can significantly help to infer/understand the physical processes at work in relativistic jets.\\
The \ct blazar is one of the brightest \gray emitters in the extragalactic sky. The source is frequently in a flaring state with the most dramatic variability being demonstrated in the \gray band. The source showed a prolonged activity in 2016-2017 when the observed daily highest flux was $(2.12\pm0.07)\times 10^{-5}\:{\rm photon\:cm^{-2}\:s^{-1}}$ corresponding to $\sim2.02\times10^{49}\:{\rm erg\:s^{-1}}$ luminosity. The \gray flux in the proper frame of the jet is $L_{\rm \gamma}=3.3\times10^{46}\:{\rm erg\:s^{-1}}$ implying that an energy much higher than $2.8\times10^{51}\:{\rm erg}$ ($>t_{\rm 1\: day}\times L_{\rm \gamma}$) should be released in the form of magnetic field and particles in order to explain the \gray emission.\\
The time-averaged \gray spectrum of \ct is best described by a log-parabola model with $\alpha=2.26$ and $\beta=0.1$ while in short time scales a substantial harder emission, $\Gamma<2.0$, with a spectrum curving at HEs is observed. Such periods were identified in the light curves with bins from 1 to 6 days. Among many periods with a hint of curvature, at least in four of them the \gray data are statistically significantly better ($>7\sigma$) described by a power-law with an exponential cut-off. All these periods were during MJD 57738-57763 when the source was in a high \gray flaring state. During these periods the cut-off energy (9.0-16.0 GeV) was relatively stable implying that it could be due to a generic feature of the process at work in the jet of \ct. This cut-off is somewhat different than that observed in the time-averaged spectrum of \ct and is clearly related with its flaring activity. External absorption can be ruled out as it is significant only for the energies above 100-200 GeV for the distance of \ct. Internal absorption cannot account for the observed curvature as well: when the emitting region is very deep inside BLR (e.g., $\sim50 r_{\rm g}$) the spectrum steepens quickly, in disagreement with the observed data, while for larger distances (e.g., $\sim1000 r_{\rm g}$) the slow drop of the flux overproduces the data observed around 10 GeV. On the other hand, the observed variability time as well as the estimated bulk Lorentz factor and the jet half opening angle put a constraint on the location of the \gray emitting region: it should be around the upper edge or outside the BLR region. The curvature observed in the \gray band is most likely due to a break/cut-off in the spectrum of radiating particles.\\
The broadband SED of \ct was modeled considering the jet dissipation occurs close (within BLR) or far from (outside BLR) the central source. The synchrotron, BLR reflected and torus photons were considered to explain the HE component in the SED of \ct. The free model parameters were estimated using the MCMC method. The observed X-ray data corresponding to the low-energy tail of the IC component limits the emitting electron maximum energy and the SSC component can reach only $1-2$ GeV, not allowing to model the observed data. When the jet plasma moves with a bulk Lorentz factor of $\Gamma=\delta=30$, the density of BLR and torus photons is comparable with or dominating over the magnetic field energy density and their IC scattering can make a significant contribution in the \gray band. Since their average energy in the jet frame exceeds that of synchrotron photons (peaking around 1 eV), the EIC component will extend beyond SSC and can explain the data above GeV. The combined SSC+EIC model can explain the observed data when the emitting electrons are distributed with a hard power-law index of $\simeq(1.7-1.8)$. On the contrary, if the jet of \ct is strongly particle dominated ($U_{\rm e}/U_{\rm B}\simeq (10^{2}-7\times10^3)$) (depressing the SSC component) the IC up-scattering of only BLR or torus photons can explain the X-ray and \gray data if the electron distribution with $\sim2.2$ index extends up to  $E_{\rm c}= 2.02 \pm 0.04$ GeV and $E_{\rm c}=3.60 \pm 0.09 $ GeV, respectively. The total jet energy ($L_{\rm jet}=L_{\rm B}+L_{\rm e}$ where $L_{B}=\pi c R_b^2 \Gamma^2 U_{B}$ and $L_{e}=\pi c R_b^2 \Gamma^2 U_{e}$) varies within $(0.04-2.3)\times 10^{47}\:{\rm erg\:s^{-1}}$ being of the same order or less than the Eddington accretion power for the black hole mass in \ct. When the \grays are produced in a separate region, the power-law index of the electrons ($2.36 \pm 0.07$) and the cut-off energy ($ 5.32 \pm 0.75$ GeV) are well constrained by the \gray data, independent of the magnetic field.\\
The estimated parameters of the electrons provided important information on the particle acceleration in the jet of \ct. The power-law index of electrons directly estimated from the X-ray or \gray data varies from $1.6$ to $2.3$- values well achievable by DSA of particles. These values cannot be directly used to put a constraint on the properties of the shock due to the complex character of the acceleration process; it can be done only under several assumptions on the unknown parameters.  However, the power-law index of electron distribution capable of explaining the data is physically realistic and it can be formed in standard relativistic shocks. On the other hand, the constraint on the cutoff of the electron distribution provides a crucial information on the diffusion of particles: from the balance of acceleration and cooling times the diffusion index should be $\alpha_{\rm diff}>2.0$ with $\eta>10^{4}$ implying that in the acceleration zone of the \ct jet the particle diffusion must be well removed from the Bohm limit ($\eta=1$ and $\alpha_{\rm diff}=1$). These parameters show that the physical environment in the jet of \ct should have a lower-level turbulence at large distances from the shocks, which results in longer diffusive mean free paths for larger momenta. These conditions are not physically unrealistic and can be formed under certain circumstances. For further discussion see \citet{2017MNRAS.464.4875B} and references therein.
\section{Conclusions} \label{sec7}
The origin of the curvature in the \gray spectra of \ct is investigated. During bright \gray flaring of \ct its emission spectrum hardened, steepening above $\sim 10$ GeV and the data are better ($>7\:\sigma$) described by a power-law with exponential cut-off model. The estimated cut-off energy remains relatively unchanged (taking into account the uncertainties).\\
The modeling of the SED of \ct allowed to constrain the free model parameters with their uncertainties, which in its turn provided information on the particle acceleration. The electron spectrum can be easily formed by diffusive shock acceleration but it is required that the diffusion occurs well beyond the Bohm limit. The prolonged \gray flaring activity of the source in 2016-2017 could be in principle due to such changes in the jet of \ct.\\
Here a single snapshot of the SED of \ct is modeled, providing a valuable information on particle acceleration and cooling processes. Observation, identification and modeling of different flaring periods characterized by a curvature in the \gray spectrum can eventually help to draw a clear picture of the global processes taking place in the blazar jets.
\section*{Acknowledgments}
This work was supported by the RA MES State Committee of Science, in the frames of the research project No 18T-1C335. This work used resources from the ASNET cloud and the EGI infrastructure with the dedicated support of CESGA (Spain).
\bibliographystyle{aa}
\bibliography{biblio}{}
\end{document}